\documentclass[aps,prl,floatfix,twocolumn,10pt,longbibliography]{revtex4-1}

\usepackage{graphicx}% Include figure files
\usepackage{dcolumn}% Align table columns on decimal point
\usepackage{bm}% Bold math
\usepackage{amssymb}% Bold math
\usepackage{rotating}% Rotate figures and tables
\usepackage[abs]{overpic}% Over pic axis
\usepackage{xcolor}% Over pic axis color
\usepackage{tabularx}% Table adjustment
\usepackage{floatrow}% Table adjustment in a row
\usepackage[caption=false]{subfig} %For side caption
\usepackage{hyperref}
\hypersetup{colorlinks = true, citebordercolor={blue}, linkcolor={blue}, citecolor={blue}, urlcolor={blue}}
\begin{document}

%\preprint{}

%\title{Signature of Off-Diagonal Long-Range Order in Exciton Condensation}

\title{Quantum Signature of Exciton Condensation}

\author{Shiva Safaei and David A. Mazziotti}

\email{damazz@uchicago.edu}

\affiliation{Department of Chemistry and The James Franck Institute, The University of Chicago, Chicago, IL 60637}%

\date{Submitted March 13, 2018; revised June 27, 2018}

% It is always \today, today,
% but any date may be explicitly specified

% \begin{abstract}
%
%   None.
%
% \end{abstract}


%Outline

\begin{abstract}
Exciton condensation, a Bose-Einstein-like condensation of excitons, was recently reported in an electronic double layer (EDL) of graphene.  We show that a universal quantum signature for exciton condensation can be used to both identity and quantify exciton condensation in molecular systems from direct calculations of the two-electron reduced density matrix.   Computed large eigenvalues in the  particle-hole reduced density matrices of pentacene and hexacene EDLs reveal the beginnings of condensation, suggesting the potential for exciton condensation in smaller scale molecular EDLs.
\end{abstract}

\maketitle

%Introduction
Exciton condensation is a Bose-Einstein-like condensation of particle-hole pairs (or excitons) into the same quantum state.  It has been realized in optical traps with polaritons~\citep{KRK2006, BKY2014, DWS2002, BHS2007}, semiconductor electronic double layers (EDLs) like gallium arsenide (GaAs)~\citep{SEP2000, KSE2002, KEP2004, TSH2004, NFE2012,DLL2017}, and most recently, EDLs of graphene~\citep{LWT2017}, proving that condensation is possible in atom-thin bilayers bound by Van der Waals forces~\citep{LTW2016, LXD2016, GGK2012, LFX2014, KFL2014, MWG2014, MBS2008, PNH2013, SM2017, KE2008, FH2016, NL2016}, as well as  the transition metal dichalcogenide $1T$-TiSe$_{2}$~\cite{Kogar2017}.  Importantly, the graphene-based experiment demonstrated that the condensation is stable upon creating an imbalance up-to 30\% in the electrons (and holes) between the two graphene layers~\cite{LWT2017}; this result reveals the potential richness of the quantum many-body states and phases associated with exciton condensation.    Like superconductivity, exciton condensation has potential applications to dissipationless energy transfer.

In this paper we examine a theoretically definitive and yet computationally practical signature of exciton condensation.  While a variety of both experimental and theoretical signatures of exciton condensation exist, they are typically not definitive indicators of exciton condensation.  The quantum signature developed here, the large eigenvalue of the (modified) particle-hole matrix~\citep{GR1969, Y1962, C1963, S1965, RM2015}, is present if and only if the fermion system exhibits condensation of particle-hole pairs (excitons) into a global state---exciton condensation.   It allows us to predict the existence and extent of exciton condensation in any quantum system from only an electronic structure calculation of the two-electron reduced density matrix (2-RDM)~\citep{M2007, M12b, M2016, M11, EJ00, N01, M04, Z04, C06, GM08, S10, M2012, V12, SHM16, P17}.   While this definition includes various phase transitions involving condensates of soft phonons that are not conventionally viewed in these terms,  such disruptive transitions, as explained by Kohn and Sherrington~\citep{Ks1970}, can be relevantly viewed in the context of exciton condensation.

The large eigenvalue of the particle-hole RDM was first presented in 1969 by Garrod and Rosina in the context of collective excitations~\citep{GR1969}.  This result provides the analogue for exciton condensation of Yang and Sasaki's general definition of fermion condensation in terms of the large eigenvalue of the 2-RDM~\citep {Y1962,C1963,S1965}.  J{\'e}rome, Rice, and Kohn~\citep{KRK1967} discussed exciton condensates, also called exciton insulators, in terms of the 2-RDM in 1967, but they did not discuss the particle-hole RDM or extend Yang and Sasaki's large eigenvalue to the particle-hole RDM.   While the paper of J{\'e}rome, Rice, and Kohn~\citep{KRK1967} and related early work~\citep{HR1968,KK1968,C1965} are well known in the study of exciton insulators, the equally important result of Garrod and Rosina~\citep{GR1969} has been largely neglected.  The large eigenvalue opens new possibilities for using electronic-structure computations to detect and study exciton condensation.  It can be used for exploring potential experimental enhancements through the tuning of chemical composition and external fields.

The paper computationally employs the large eigenvalue to show evidence for the beginnings of exciton condensation in pentacene and hexacene EDLs.  This result provides evidence for the formation of an exciton condensate in a molecular-scale system without the application of external fields.  Varsano {\em et al.}~\citep{VSS2017} recently predicted the formation of an exciton insulator (condensate) in carbon nanotubes, sharing some of the characteristics of the acene EDLs.  Non-trivial direct calculations of the two-electron reduced density matrix (2-RDM) capture strong electron correlation including the long-range order of the exciton condensation.  Such calculations, cited by the National Research Council in 1995 as one of the top outstanding problems in physical science, have only become possible recently~\citep{M2007, M12b, M2016, M11, EJ00, N01, M04, Z04, C06, GM08, S10, M2012, V12, SHM16, P17} and correspond to wavefunction calculations where the wavefunction, if constructible as a combination of determinants, would require a billion times the degrees of freedom treatable today by state-of-the-art supercomputers.  Molecular-scale condensates, screened by electronic structure calculations, may be useful for the development of dissipationless molecular circuits and devices, especially in the context of molecular electronics.

\begin{figure}
\centering
\sidesubfloat[]{\label{fig:H_chain}
 \begin{overpic}[trim={7.5cm 19.8cm 4cm 22cm},clip,scale=0.29]{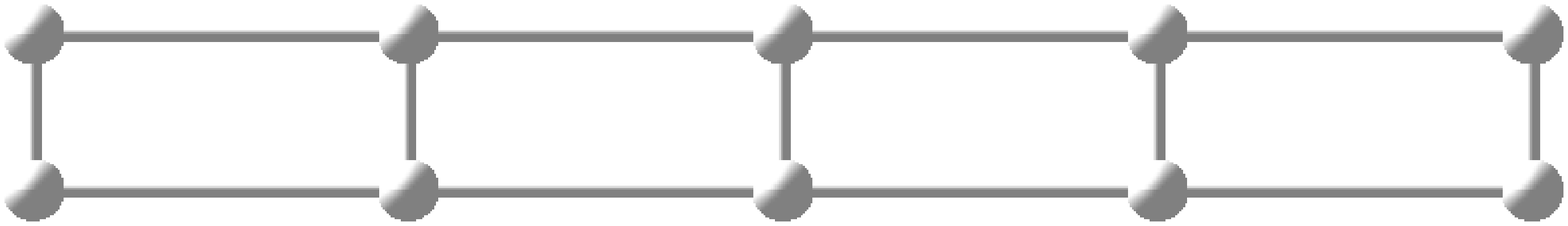}
       \put(-5,10){\color{gray}\vector(1,0){15}}\put(11,9){\color{gray}$x$}
       \put(-5,10){\color{gray}\vector(0,1){15}}\put(-5,27){\color{gray}$z$}
       \put(-5,10){\color{gray}\vector(-1,-1){10}}\put(-20,-1){\color{gray}$y$}
       \put(90,23){\color{red}\vector(1,0){26}}\put(90,23){\color{red}\vector(-1,0){26}}
       \put(72,15){\color{red}\footnotesize{$\ell=6.0$ \AA}}
       \put(122,35){\color{red}\vector(0,1){14}}\put(122,35){\color{red}\vector(0,-1){9}}
       \put(124,35){\color{red}\footnotesize{$d=2.5$ \AA}}
 \end{overpic}}\\
\vspace{5mm}
\sidesubfloat[]{\label{fig:H_eval}\includegraphics[scale=.4]{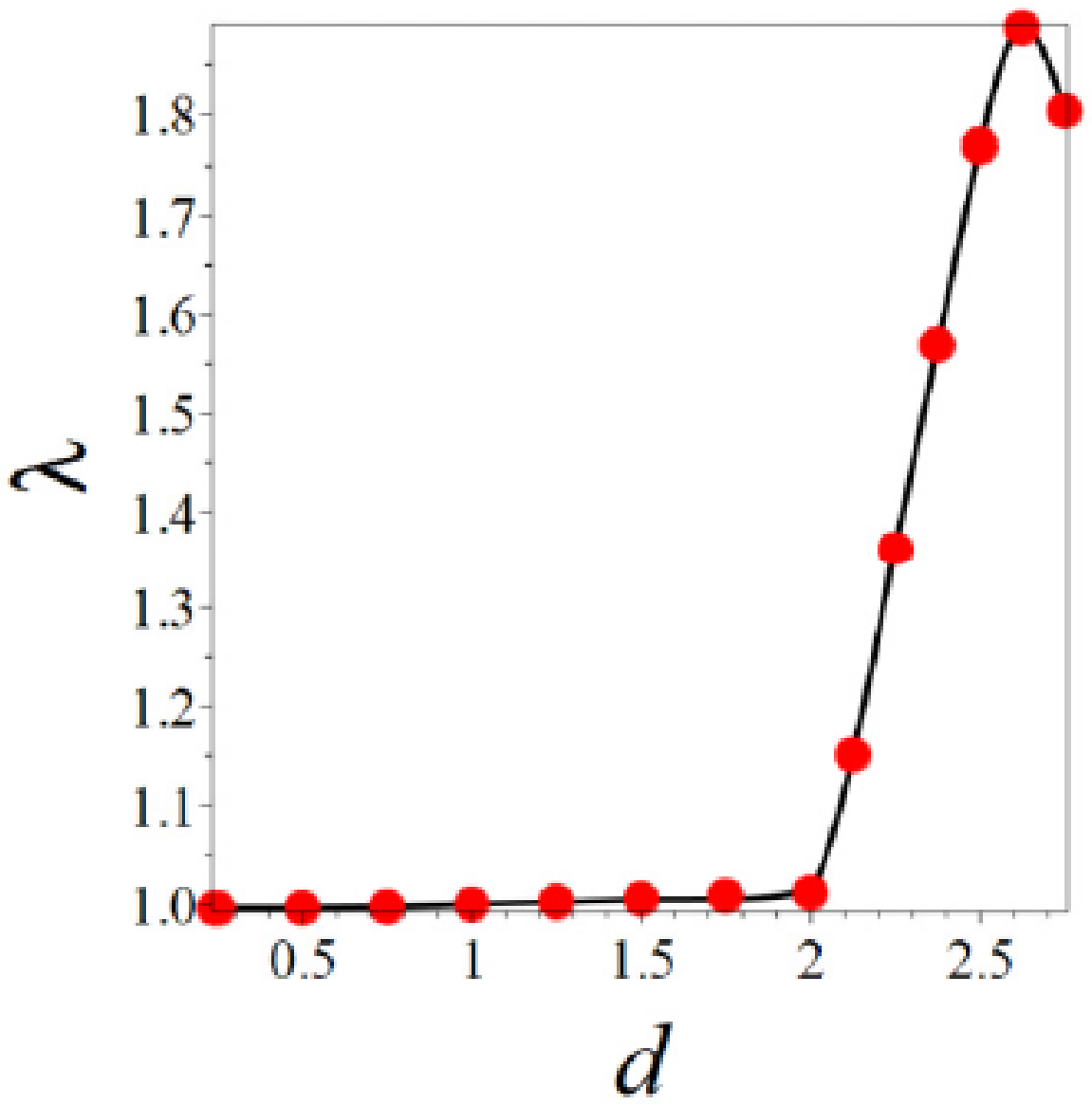}}
\caption{(a) A schematic of the hydrogen-chain EDL shows the distance $\ell$ between hydrogen atoms in each chain and the distance $d$ between the two chains. (b) The largest eigenvalue of the modified particle-hole RDM is shown as a function of the parameter $d$ in \AA\ for $\ell = 6$~\AA.}
\label{fig:H_1}
\end{figure}

%%% Figure %%%

\begin{figure}[ht!]
  \centering
  \sidesubfloat[]{\label{fig:H5_1}\includegraphics[scale=0.4]{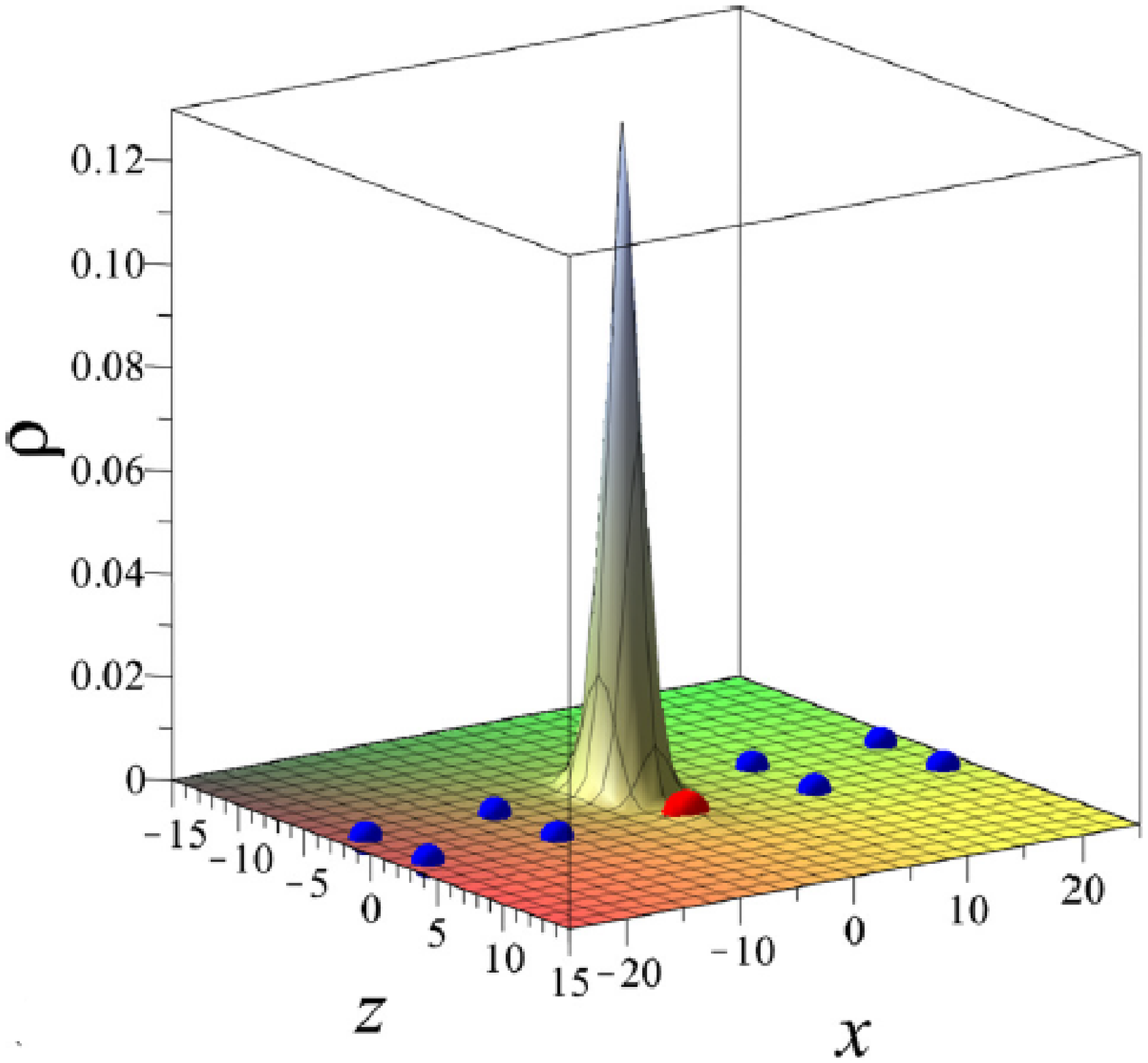}}\\
  \sidesubfloat[]{\label{fig:H5_2}\includegraphics[scale=0.4]{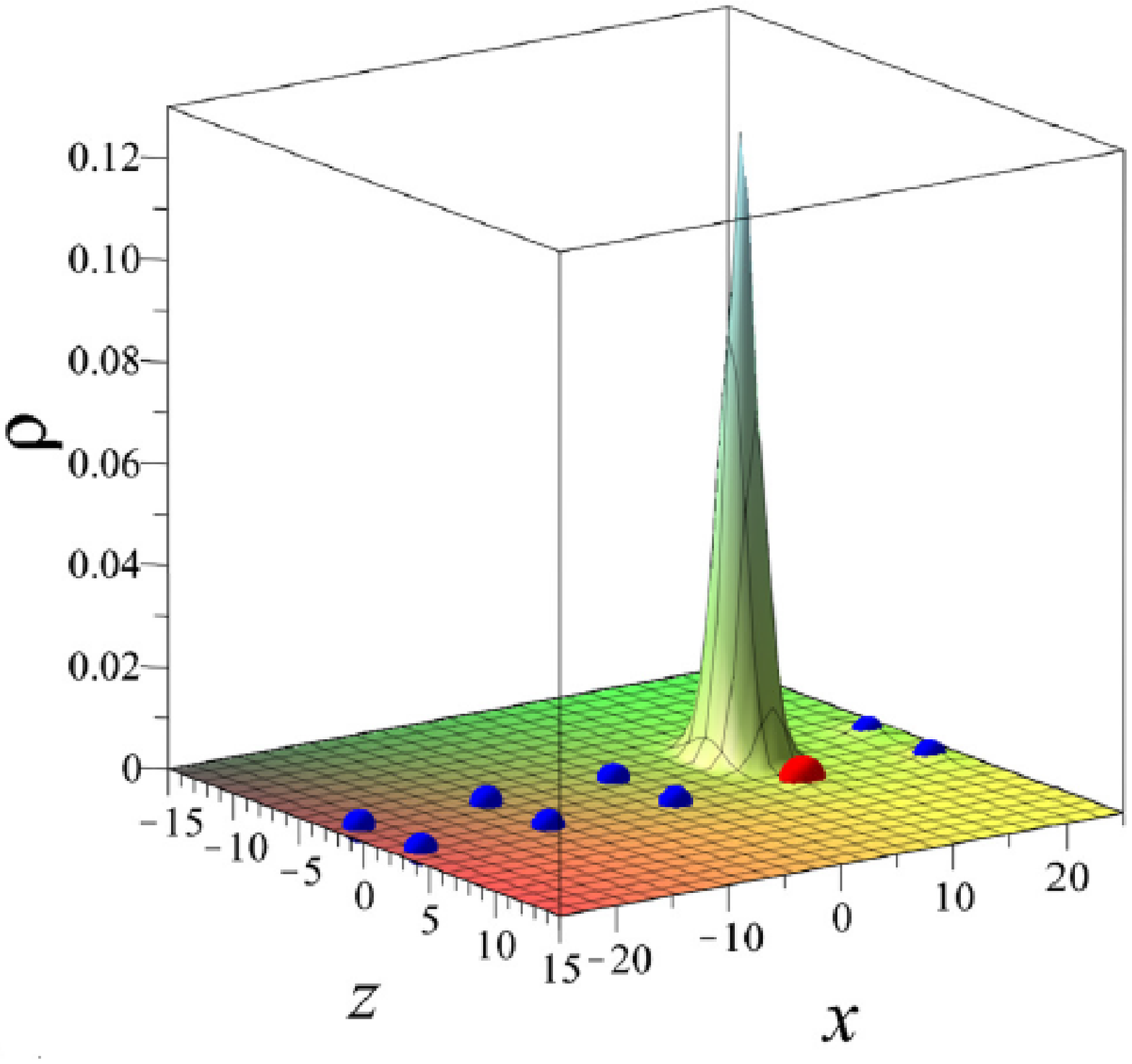}}\\
  \sidesubfloat[]{\label{fig:H5_3}\includegraphics[scale=0.4]{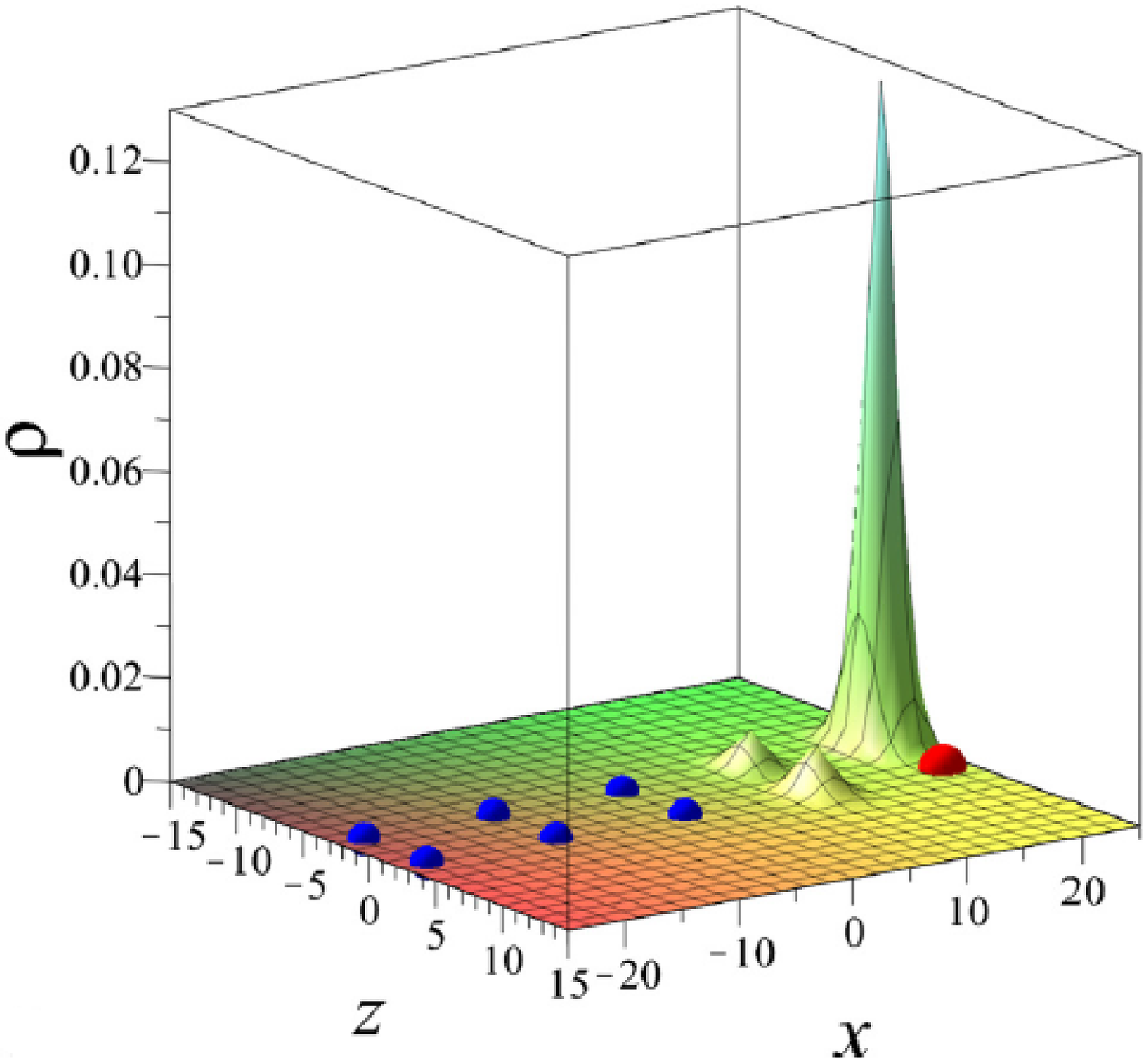}}
  \caption{For the hydrogen-chain EDL the particle-hole pairing in the eigenstate associated with the largest eigenvalue of the modified particle-hole RDM is shown. In (a), (b), and (c) the probability density of the hole is shown for the particle placed at the position of the red dot.  Coordinates $z$ and $x$ are shown in atomic units (a.u.) where 1~a.u. = 0.529~\AA.} \label{fig:H_2}
\end{figure}

\begin{figure}[ht!]
\centering
\sidesubfloat[]{\label{fig:Hex_chain}
  \begin{overpic}[trim={13cm 15cm 9cm 16cm},clip,scale=0.28]{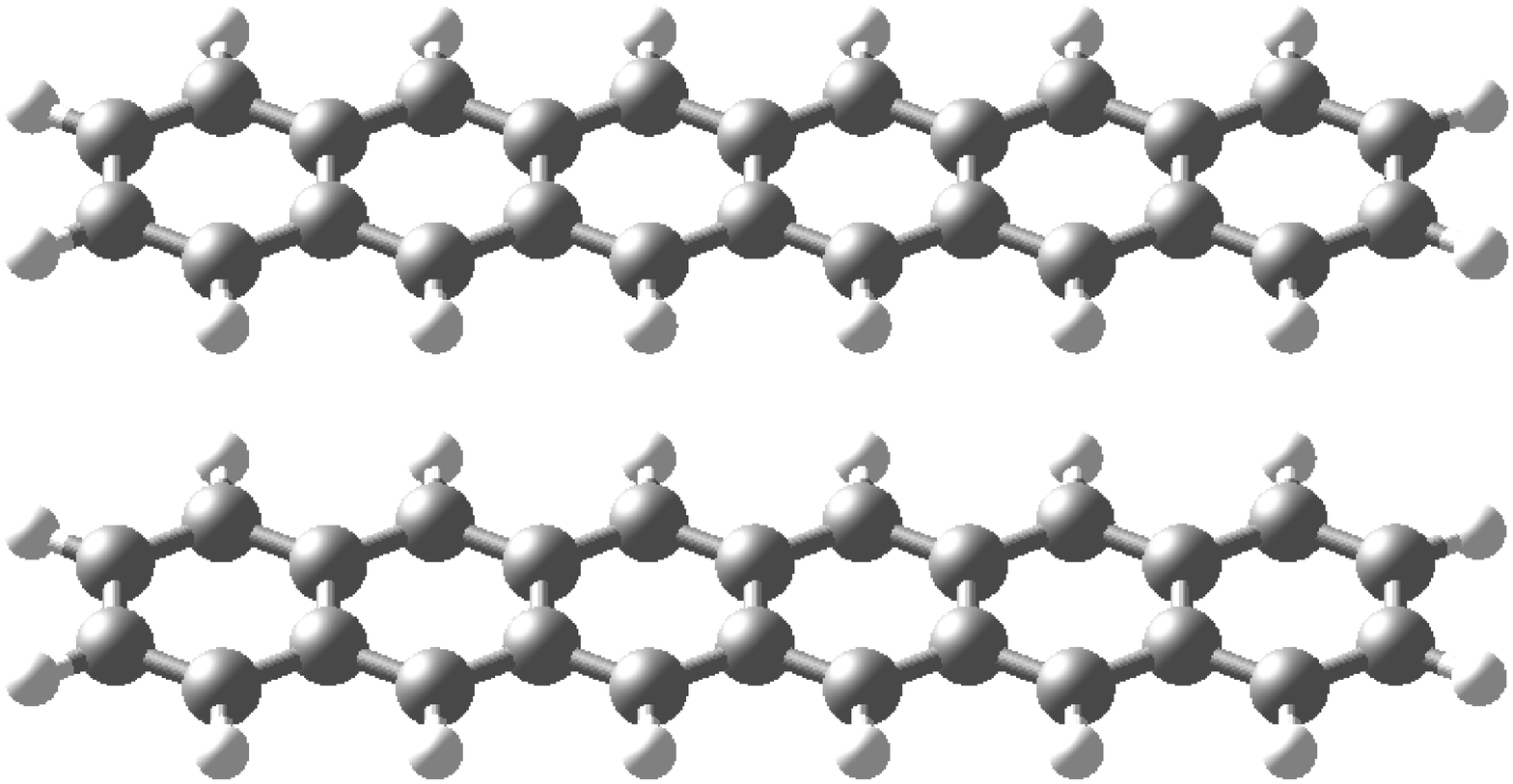}
     \put(-5,60){\color{gray}\vector(1,0){15}}\put(11,59){\color{gray}$x$}
     \put(-5,60){\color{gray}\vector(0,1){15}}\put(-5,77){\color{gray}$z$}
     \put(-5,60){\color{gray}\vector(-1,-1){10}}\put(-20,49){\color{gray}$y$}
     \put(123,85.5){\color{red}\vector(0,1){9}}\put(123,67.5){\color{red}\vector(0,-1){37}}
     \put(125,59){\color{red}\footnotesize{$d=2.5$ \AA}}
     \end{overpic}}\\
\vspace{5mm}
\sidesubfloat[]{\label{fig:Hex_eval}\includegraphics[scale=.4]{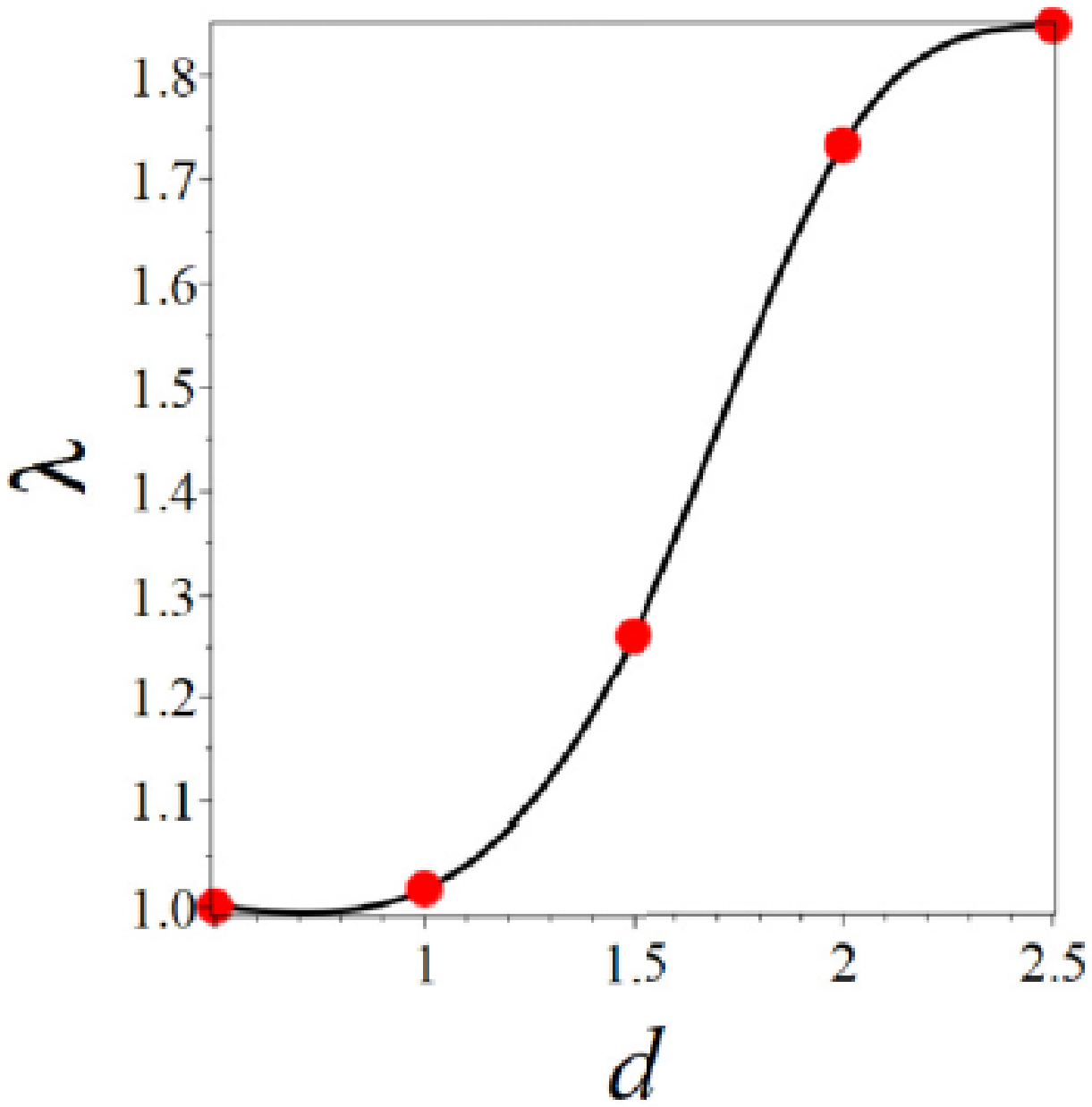}}
\caption{(a) A schematic picture of the hexacene EDL is shown. (b) The largest eigenvalue of the modified particle-hole RDM of the hexacene EDL reaches a maximum of 1.8 around $d$ = 2.5~\AA.}
\label{fig:Hex_1}
\end{figure}

%%% Figure %%%

\begin{figure}[ht!]
  \centering
   \sidesubfloat[]{\label{fig:Hexacene_1}\includegraphics[scale=0.407]{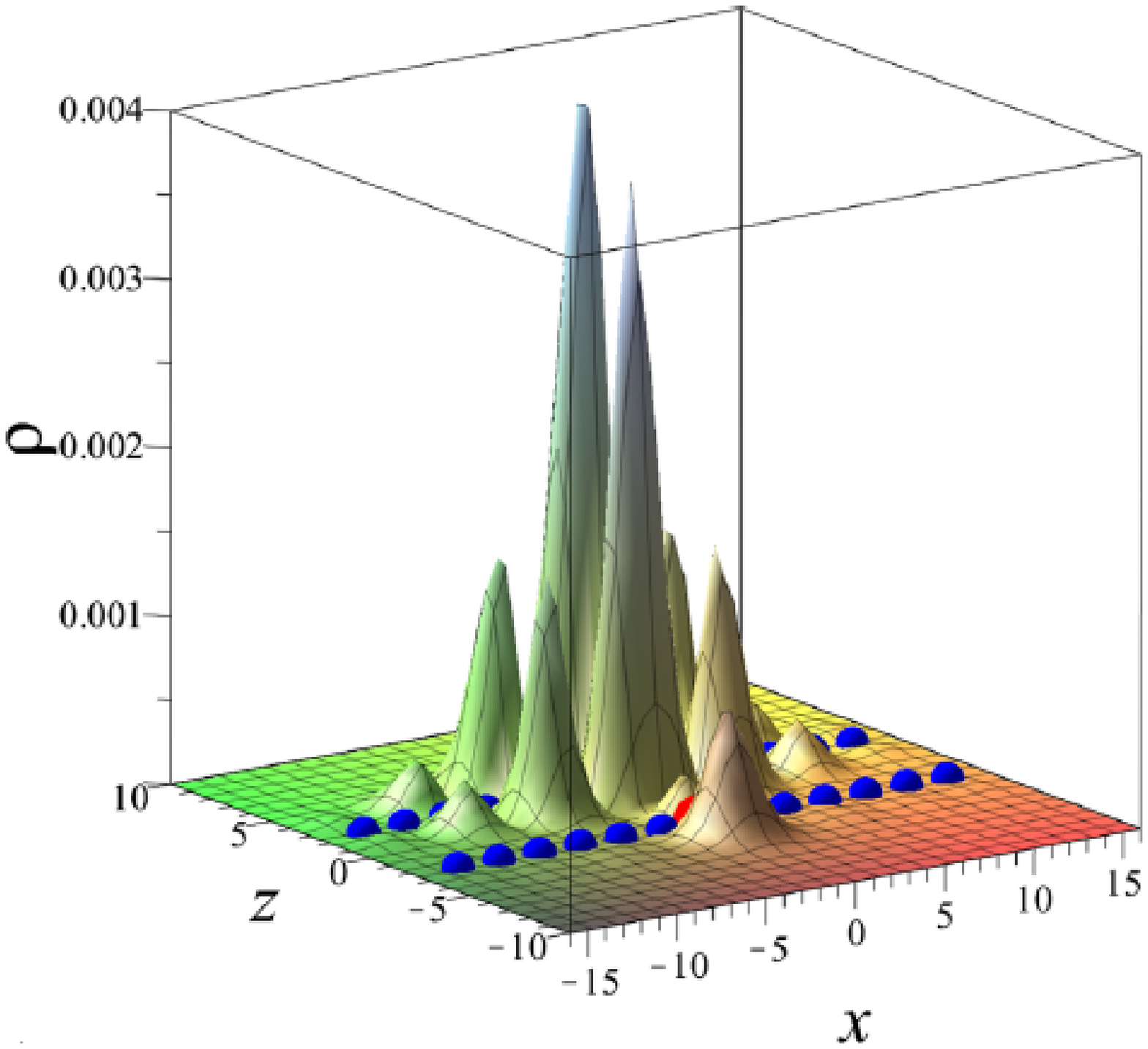}}\\
   \sidesubfloat[]{\label{fig:Hexacene_2}\includegraphics[scale=0.38]{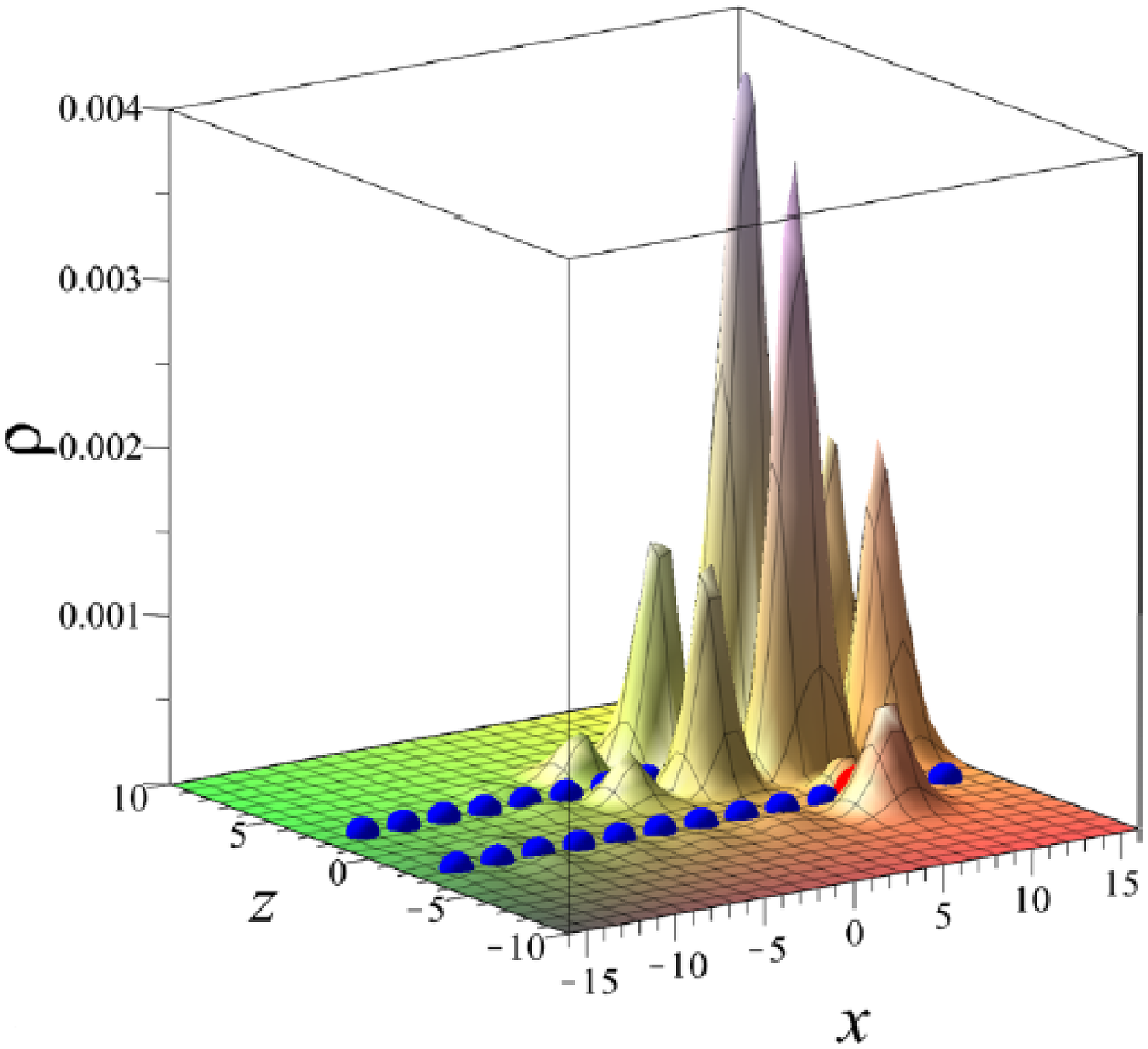}}\\
   \sidesubfloat[]{\label{fig:Hexacene_3}\includegraphics[scale=0.38]{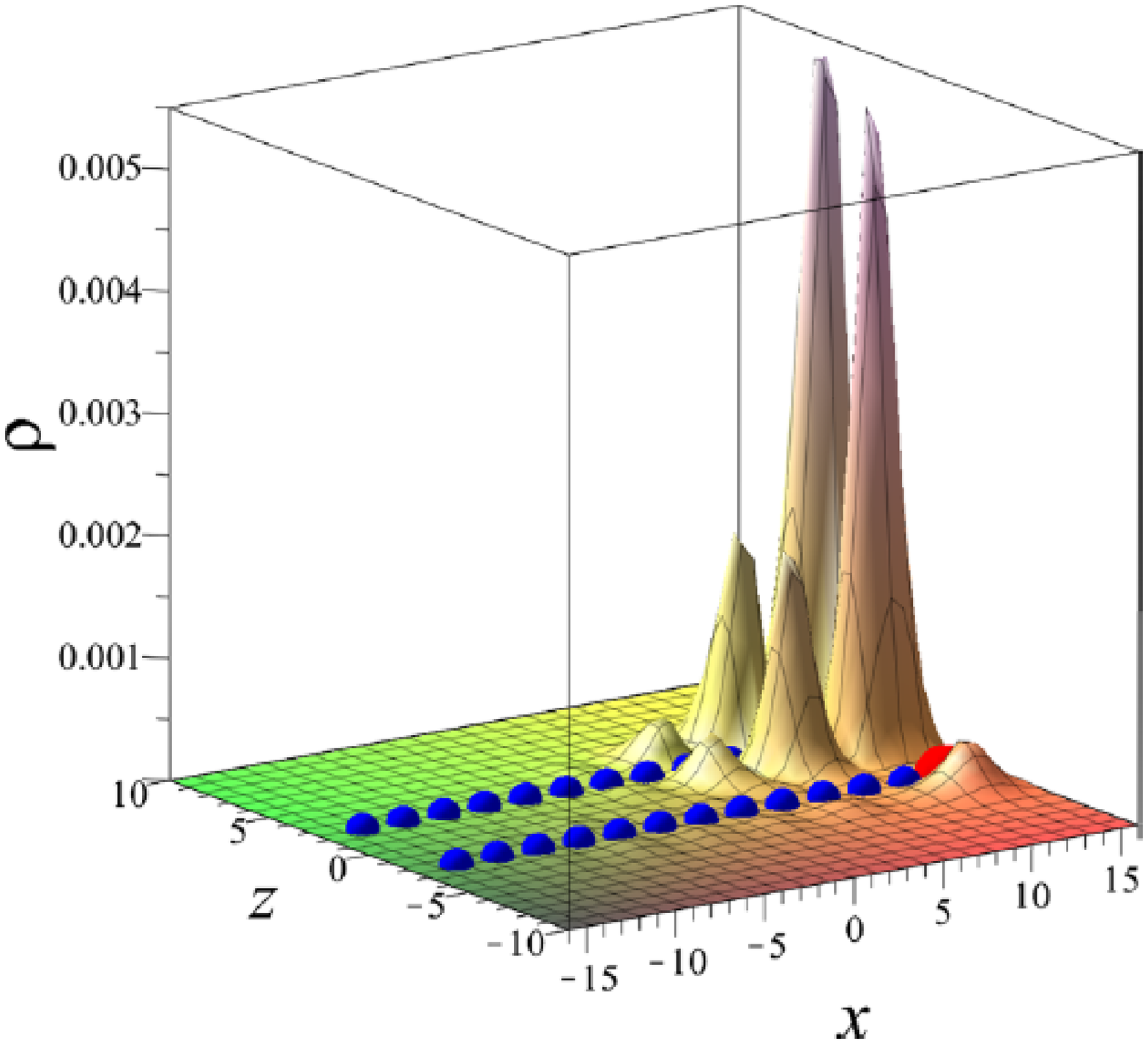}}
  \caption{For the hexacene EDL the particle-hole pairing in the eigenstate associated with the largest eigenvalue of the modified particle-hole RDM is shown. In each graph the probability density of the hole is shown for the particle placed on the layer at $z$ = -2.36~a.u. at the position of the red dot.  In (a), (b), and (c) the red dot is center, right of center, and far right of center, respectively.  Coordinates $z$ and $x$ are shown in atomic units (a.u.) where 1~a.u. = 0.529~\AA.}\label{fig:Hex_2}
\end{figure}

In Bose-Einstein condensation bosons condense upon cooling into the same lowest-energy orbital.  For example, in the alkali-metal Bose-Einstein experiments~\citep{AEM1995,DMA1995,BST1995} nearly a mole of alkali-metal bosons condense into the lowest Gaussian orbital of a harmonic magnetic trap.  Signature of Bose-Einstein condensation is a large eigenvalue in the one-boson RDM given by
\begin{equation}
^{1} D(1;{\bar 1}) = \langle \Psi | {\hat \psi}^{\dagger}(1) {\hat \psi}({\bar 1}) | \Psi \rangle
\end{equation}
where $\Psi$ is the $N$-boson wave function, each roman number represents the spatial coordinates and spin component of a boson, and the quantum-field operators ${\hat \psi}^{\dagger}(1)$ and ${\hat \psi}({\bar 1})$ create or annihilate a boson~\citep{Y1962,MH1999}. Unlike either fermion or exciton condensation Bose-Einstein condensation is not driven by electron correlation; the wave function of a noninteracting pure-state Bose-Einstein condensate is a product of the single orbital $\phi(1)$ in which the bosons condense.

In fermion condensation pairs of fermions condense into a single two-electron function known as a geminal~\citep{Y1962,C1963,S1965,RM2015}. In superconductivity, for example, electrons pair to form Cooper pairs that condense into a single geminal with long-range order~\citep{BCS1975}. As shown by Yang~\citep{Y1962} and Sasaki~\citep{C1963,S1965} independently, the signature of a fermion condensation is a large eigenvalue in the 2-RDM given by
\begin{equation}
^{2} D(12;{\bar 1}{\bar 2}) = \langle \Psi | {\hat \psi}^{\dagger}(1) {\hat \psi}^{\dagger}(2) {\hat \psi}({\bar 2}) {\hat \psi}({\bar 1})  | \Psi \rangle
\end{equation}
where $\Psi$ is the $N$-fermion wavefunction and the quantum-field operators $ {\hat \psi}^{\dagger}(1)$ and  ${\hat \psi}(1)$ create and annihilate a {\em fermion} at position~1. Even though Pauli exclusion principle shows that the maximum fermion occupation of an orbital is bounded from above by one, Yang~\citep {Y1962} and Sasaki~\citep{C1963,S1965} proved that the maximum eigenvalue of the 2-RDM can be proportional to the number $N$ of fermions in the limit of strong correlation.

Exciton condensation, distinct from both Bose-Einstein and fermion condensations, is the condensation of particle-hole pairs (excitons) into a single particle-hole function.  By analogy with the well-known cases of Bose-Einstein and fermion condensations, the signature of exciton condensation is two large eigenvalues in the particle-hole RDM~\citep{GR1969} given by
\begin{equation}
^{2} G(1{\bar 1};2{\bar 2}) =  \langle \Psi | {\hat \psi}^{\dagger}(1) {\hat \psi}({\bar 1}) {\hat \psi}^{\dagger}({\bar 2}){\hat \psi}(2)  | \Psi \rangle
\end{equation}
where $\Psi$ is the $N$-fermion wavefunction and $ {\hat \psi}^{\dagger}(1)$ and  ${\hat \psi}(1)$ are fermion quantum-field operators.  There are two large eigenvalues in the case of exciton condensation because the particle-hole RDM, even in the noninteracting limit, always has one large eigenvalue corresponding to a ground-state-to-ground-state projection rather than an excitation. This eigenvalue, which might be viewed as spurious as it is unrelated to exciton condensation, can be removed by subtraction of its ground-state resolution to generate a modified particle-hole matrix, having a single large eigenvalue upon condensation
\begin{equation}
^{2} {\tilde G}(1{\bar 1};2{\bar 2}) = ^{2} G(1{\bar 1};2{\bar 2}) - ^{1} D(1;{\bar 1}) ^{1} D(2;{\bar 2}) .
\end{equation}
In the noninteracting limit all of the eigenvalues of the modified particle-hole RDM are equal to zero or one; furthermore, in typical molecular systems the largest eigenvalue is very nearly one.  In early work on reduced density matrices Garrod and Rosina showed that the largest eigenvalue of the modified particle-hole matrix, or the maximum number of excitons in the condensate, is bounded from above by $N/2$~\citep{GR1969}.

Formation of the exciton condensate requires particle-hole pairs that can occupy the same particle-hole function.  This type of pairing can be achieved through spatial symmetry~\citep{KK1965,LY1976,EM2004}. Exciton condensation has been achieved in EDLs of GaAs~\citep{SEP2000,KSE2002,KEP2004,TSH2004,NFE2012} and graphene~\citep{LTW2016,LXD2016,GGK2012,LFX2014,KFL2014,MWG2014,LWT2017}. The double layers allow the particles in one layer to pair with the holes in the opposite layer and vice versa. Conceptually, we can picture a checkerboard of particles (black) and holes (white) in one layer paired with an inverted checkerboard of holes (white) and particles (black) in the other layer~\citep{EM2004}.  Such a pairing requires that the particles and holes become strongly entangled  In contrast, Bose-Einstein condensation requires neither pairing nor entangling of the bosons.  As a consequence, exciton condensation has more in common with fermion condensation where the fermions form a strongly correlated Cooper pair that participates in the condensation.

Nonetheless, fermions exploit a different symmetry to achieve pairing than the particles and holes which pair to form excitons. Due to their exchange symmetry, fermions occupy a two-fermion function $g(12)$ that is antisymmetric in the exchange of the coordinates of particles 1 and 2, that is $g(21) = - g(12)$.  The function $g(12)$ can be viewed as an antisymmetric matrix whose trace is zero and whose imaginary eigenvalues are paired~\citep{CY00}:
\begin{equation}
\int{g(12) \phi_{\pm i} (2) d2} = \pm \epsilon_{i} \phi_{\pm i}(1) .
\end{equation}
 Although this pairing is present in all systems of fermions, it can lead to fermion condensation when the occupations $\epsilon_{i}$ become degenerate for multiple pairs of orbitals, i.e. $\phi_{+i}$ and $\phi_{-i}$ for a range of $i$.  Fermion condensation, whether induced by phonons as in BCS superconductivity~\cite{BCS1975} or another mechanism, exploits the antisymmetry of fermions to support the condensation of multiple fermion pairs into the same two-fermion state $g(12)$.  In contrast, the particle-hole function $f(12)$ occupied by excitons in exciton condensation lacks an intrinsic pairing due to antisymmetry.  The pairing of the particle and the hole must be accomplished by another symmetry such as the spatial symmetry created by the electron double layer (EDL).  The EDL symmetry satisfies the fundamental symmetry requirement for the pairing of particles and holes, and consequently, molecules and materials in EDL formation are likely candidates for exciton condensation.
\nopagebreak
%Results

Using the large eigenvalue of the particle-hole RDM as a quantum signature of the exciton condensation, we examined two molecular-scale bilayers: (1) a stretched hydrogen-chain EDL and (2) face-to-face pentacene and hexacene EDLs.  For each system the 2-RDM was directly determined without computation of the many-electron wave function. The energy was computed as a variational functional of the 2-RDM~\citep{M2007, M12b, M2016, M11, EJ00, N01, M04, Z04, C06, GM08, S10, M2012, V12, SHM16, P17,VB1996} which was constrained by $N$-representability conditions~\citep{M2016, M2012, P17, SGC13} that are necessary for it to represent an $N$-electron quantum system.  Because the $N$-representability conditions are non-perturbative with polynomial computational scaling, the variational 2-RDM method is able to catpture strong correlation in molecules that are too correlated to treat by density functional theory or single-reference {\em ab initio} methods and yet too large to treat by either full or partial configuration-interaction calculations.  Applications have recently been made to studying strong correlation in transition-metal complexes~\citep{SHM16} and the nitrogenase catalyst FeMoco~\citep{MM2018} as well as computing molecular conductivity in benzenedithiol~\citep{SM2018}.  Finally, the particle-hole RDM, containing the signature of exciton condensation, was computed by a linear mapping of the 2-RDM. Additional details are provided under Methods in the Supplemental Information~\citep{SM2018b}.
\nopagebreak

A sketch of the hydrogen-chain EDL is shown in Fig.~\ref{fig:H_chain}  As shown, each chain was chosen to have 5 hydrogen atoms. Two geometric parameters in the EDL are the distance $\ell$ between hydrogen atoms in each chain and the distance $d$ between the two chains.  The two-dimensional parameter space defined by $d$ and $\ell$ was explored to maximize the largest eigenvalue of the particle-hole RDM.  We found that the optimal $\ell$ is approximately 6~\AA\, corresponding to a highly stretched geometry. With $\ell = 6$~\AA\ the plot in Fig.~\ref{fig:H_eval} shows the largest eigenvalue of the modified particle-hole RDM as a function of the parameter $d$.  As $d$ increases from 2~\AA\ to 2.6~\AA\, we observe a rapid increase in the largest eigenvalue indicating the formation of a condensate with approximately two excitons. For $d$ less than 2~\AA\  covalent bonds form between the layers, quenching the possibility for condensation as electrons pair in the bonding orbital. Furthermore, for $d$ greater than a certain distance (3~\AA\ in this case) the particles and holes of the two layers become too independent to form excitons.  Table~1 in the Supplemental Information~\citep{SM2018b} also shows a second large eigenvalue in the modified particle-hole RDM, indicating some exciton condensate in a second particle-hole eigenstate.  The second large eigenvalue is important because it reveals the potential richness of possible condensate states and phases.  Table~2 in the Supplemental Information~\citep{SM2018b} further shows that the magnitude of the largest eigenvalue generally increases with the length of the hydrogen EDLs.
\nopagebreak
For the hydrogen-chain EDL Fig.~\ref{fig:H_2} shows the particle-hole pairing in the eigenstate associated with the largest eigenvalue of the modified particle-hole RDM. In each panel the probability density of the hole is shown for the particle placed at the position of the red dot.  The probability of finding a particle on a hydrogen atom in one layer, we observe, is paired with the probability of finding the hole on the opposite hydrogen atom in the other layer. This pairing occurs at each site along the chain, revealing the off-diagonal long-range order of the exciton condensate.

A schematic picture of the hexacene EDL is shown in Fig.~\ref{fig:Hex_chain}. The two hexacene chains, each consisting of six fused benzene rings, are stacked on top of each other to form a face-to-face EDL, providing a molecular quasi-analog of the graphene EDL.  While the hydrogen-chain EDL has two parameters $d$ and $\ell$, the hexacene EDL has only one adjustable parameter $d$, the distance between the layers. For the hexacene EDL Fig.~\ref{fig:Hex_eval} shows the largest eigenvalue of the modified particle-hole RDM as a function of the distance $d$.  The largest eigenvalue begins to increase from one around 1~\AA\, reaching a maximum value of approximately 1.8 around 2.5~\AA\ before decreasing as $d$ increases further. As in the case of the hydrogen-chain EDL, there is a sweet spot in the distance $d$ at which the chains are sufficiently separated to prevent bonding and yet sufficiently close to enable particle-hole entanglement to form the exciton.   Figure~1 in the Supplemental Information~\citep{SM2018b} shows a similar plot of the largest eigenvalue for a pentacene EDL with a structure similar to the hexacene EDL.  We observe that the largest eigenvalue peaks at 1.6 around 2.5~\AA\, indicating that the degree of exciton condensation is increasing with increasing chain length.  The largest eigenvalue of the benzene EDL is only 1.195 at a separation of 2.5~\AA\.  Table~3 in the Supplemental Information~\citep{SM2018b} reveals that, as in the case of the hydrogen-chain EDL, some exciton condensate forms in the second and third largest eigenvalues of the modified particle-hole matrix for both the pentacene and hexacene EDLs.

For the hexacene EDL Fig.~\ref{fig:Hex_2} shows the particle-hole pairing in the eigenstate associated with the largest eigenvalue of the modified particle-hole RDM.  As for the hydrogen-chain in Fig.~\ref{fig:H_2} the probability density of the hole is displayed with the red dot representing the position of the particle.  Significantly, the Dirac-delta probability distribution of a particle on a benzene ring in one layer is paired with a probability distribution of a hole that is highly localized on the opposite benzene ring of the adjacent layer.  The three panels reveal that this pairing of a particle and a hole is present at all sites along the chain, indicating the presence of off-diagonal long-range order in the eigenstate associated with the largest eigenvalue. Analogous plots for the eigenstates of the modified particle-hole RDM that are not associated with exciton condensation show a highly delocalized hole probability density that is not consistent with condensation.

%Conclusions

The large eigenvalue of the modified particle-hole RDM is a definitive, universal quantum signature of exciton condensation. Unlike the pairing of fermions in superconductivity that is driven by anti-symmetry, the pairing of particles and holes in exciton condensation is driven by a more system-specific symmetry such as the geometric symmetry in the EDLs of GaAs~\citep{SEP2000,KSE2002,KEP2004,TSH2004,NFE2012} and graphene~\citep{LTW2016, LXD2016, GGK2012, LFX2014, KFL2014, MWG2014, MBS2008, PNH2013, SM2017, KE2008, FH2016, NL2016}.  Computations of molecular EDLs including a stretched hydrogen-chain EDL as well as pentacene and hexacene EDLs reveal large eigenvalues of the particle-hole matrix for a range of suitable interlayer distances. These results may indicate that exciton condensation is experimentally realizable in smaller scale molecular EDLs and that it is present in a potentially richer array of systems, states, and phases than previously conjectured.  Such potential richness, also observed in the results of recent graphene-EDL experiments, is critical to achieving practical applications of exciton condensation such as dissipation-free energy transfer.

\begin{acknowledgments}

D.A.M. gratefully acknowledges the United States-Army Research Office (ARO) Grants W911NF-16-C-0030 and W911NF-16-1-0152, the United States National Science Foundation Grant CHE-1565638, and the United States Air Force Office of Scientific Research Grant FA9550-14-1-0367.  D. A. M. also thanks D. R. Herschbach, H. A. Rabitz, and A. R. Mazziotti for their encouragement and support.

\end{acknowledgments}

%\bibliography{referencesR6}

%

\end{document}